# Quantitative Analysis of Desirability in User Experience

**Sisira Adikari**
School of Information Systems and Accounting
University of Canberra
Canberra, Australia ACT 2617
Email: sisira.adikari@canberra.edu.au

**Craig McDonald**
School of Information Systems and Accounting
University of Canberra
Canberra, Australia ACT 2617
Email: craig.mcdonal@canberra.edu.au

**John Campbell**
School of Information Systems and Accounting
University of Canberra
Canberra, Australia ACT 2617
Email: john.campbell@canberra.edu.au

## Abstract

The multi-dimensional nature of user experience warrants a rigorous assessment of the interactive experience in systems. User experience assessments are based on product evaluations and subsequent analysis of the collected data using quantitative and qualitative techniques. The quality of user experience assessments is dependent on the effectiveness of the techniques deployed. This paper presents the results of a quantitative analysis of desirability aspects of the user experience in a comparative product evaluation study. The data collection was conducted using 118 item Microsoft Product Reaction Cards (PRC) tool followed by data analysis based on the Surface Measure of Overall Performance (SMOP) approach. The results of this study suggest that the incorporation of SMOP as an approach for PRC data analysis derive conclusive evidence of desirability in user experience. The significance of the paper is that it presents a novel analysis method incorporating product reaction cards and surface measure of overall performance approach for an effective quantitative analysis which can be used in academic research and industrial practice.

**Keywords** User Experience, Product Reaction Cards, Surface Measure of Overall Performance

## 1　Introduction

There are many definitions of user experience from the literature. Hassenzahl and Tractinsky (2006) consider user experience as ranging from traditional usability to beauty, hedonic, affective or experiential aspects of technology use. According to Kuniavsky (2010, p.10), user experience is the totality of user perceptions associated with their interactions with an artefact (product, system, or service) in terms of effectiveness (how good is the result?), efficiency (how fast it?), emotional satisfaction (how good does it feel?), and the quality of the relationship with the entity that created the artefact. The ISO standard 9241-210 (2009) defines user experience as perceptions of persons and responses that result from the use or anticipated use of a product, system or service emphasising two main aspects: use and anticipated use. The ISO standard also points out that user experience includes all the users' emotions, beliefs, preferences, perceptions, physical and psychological responses, behaviours and accomplishments that occur before, during and after use, and usability criteria can be used to assess aspects of user experience (ISO, 2009). These two definitions by Kuniavsky (2010) and ISO standard 9241-210 show the complex and multifaceted view of user experience. According to these definitions, user experience assessment needs to be an evaluation of the total users' interactive experience of a product, system or service. As highlighted by Adikari et al. (2011), the interactive experience is the combined result of use (i.e. actual interaction experience), anticipated use (i.e. pre-interaction experience such as needs and expectations), and after use (post-interaction experience), and these three components are equally important for consideration in user experience assessments.

Based on the analysis of the ISO standard 9241-210, Bevan (2009) considers that if user experience includes all behaviour, it presumably includes users' effectiveness and efficiency and it seems consistent with the methods nominated by many people in industry who appear to have subsumed usability within user experience. As defined in ISO standard 9241-11, usability is concerned with the effectiveness, efficiency and satisfaction with which specified users achieve specified goals in particular environments. Bevan's view indicates that user experience is not distinct, and it is an extension of usability. Preece et al. have explained this broader view of usability within user experience (2002, p. 19) in terms of user experience goals and usability goals emphasising that user experience is at a level beyond that of usability. According to them, user experience occurs as a result of achieving usability goals during an interaction (see Figure 1). Moreover, they point out that user experience goals are more concerned with how users experience an interactive system from their perspective rather than assessing how useful or productive a system is from the product's own perspective. Their definition of user experience goals is: satisfying, enjoyable, fun, entertaining, helpful, motivating, aesthetically pleasing, supportive of creativity, rewarding and emotionally fulfilling.

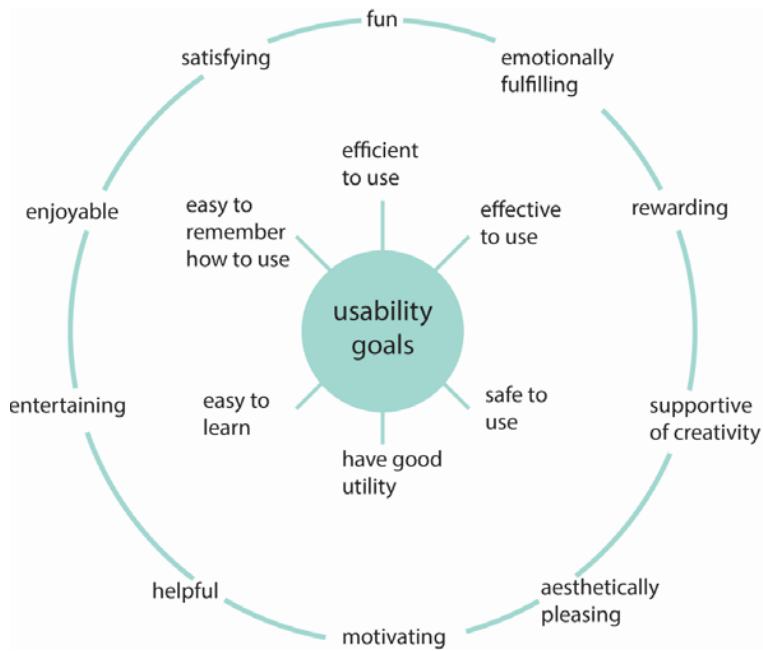

*Figure 1: Usability and UX goals (Preece et al., 2002, p. 19)*

As shown in Figure 1, their model of usability and user experience consists of six usability goals: efficient to use, effective to use, safe to use, having good utility, easy to learn, and easy to remember. Importantly, the model does not consider 'satisfaction' as a usability goal at the operational level; instead, it shows as a user experience goal. Another important difference in this model is that of 'safety' which has been included as a primary usability goal. Later, the same authors (Rogers et al. 2011, pp. 23-25) presented an updated model of user experience goals in two dimensions; desirable (positive) ones, and undesirable (negative) ones. The desirable user experience goals are: satisfying, enjoyable, engaging, pleasurable, exciting, entertaining, helpful, motivating, challenging, enhancing sociability, supporting creativity, cognitively stimulating, fun, provocative, surprising, rewarding, and emotionally fulfilling. The undesirable user experience goals are: boring, frustrating, making one feel guilty, annoying, childish, unpleasant, patronizing, making one feel stupidity, cutesy and gimmicky. Rogers et al. (2011, pp. 19-21) also pointed out that interaction design should not only set usability goals for product design but also set user experience goals to assess whether the product is enjoyable, satisfying and motivating. They described many of these goals as subjective qualities concerned with how a system feels to a user. They also highlight that not all usability and user experience goals will be relevant to the design and evaluation of an interactive product (or system) as some combinations will not be compatible.

Accordingly, effective user experience assessments should include evaluation criteria that consider usability goals as well as user experience goals, which are mostly subjective qualities that are experienced by the user during an interaction. Some of the user experience evaluation criteria indicate interrelation among other criteria. For example, 'enjoyable (joy)' is most likely to be dependent on 'fun', and vice versa. These factors highlight the emotional impact on the user during an interaction and point out the desirability of an artefact from the user's point of view. There are many user experience evaluation methods used in academic and industrial contexts under the categories of lab studies, field studies, surveys, and expert evaluations (Roto et al., 2009). These methods explore different attributes of user experience including some elements of desirability. Desirability in user experience and its' importance have been acknowledged (Hartson and Pyla, 2012, p.32). Recent studies have reported that Product Reaction Cards (PRC) are more effective at expanding the understanding of user experience and desirability (Barnum and Palmer, 2010A, 2010B).

This paper presents the results of a quantitative analysis of user experience of a comparative product evaluation study. Two software products were developed by two distinct agile software development teams based on the same suite of user stories (requirements) presented by a product owner. One product development team followed an integrated agile and user experience design approach and developed the product designated as EAP product, while the other team developed the product designated as CAP product based on traditional agile software development approach. The integrated approach intended to provide only required details of user experience related information as needed to support the analysis and design during agile software development iterations.

## 2  Product Reaction Cards for Data Collection

Product reaction cards (Benedek & Miner, 2002) aim to elicit the aspects of perceived desirability from a user on the experience of a product. The desirability of user experience is the least tangible aspect of a good user experience, and is concerned with aspects such as imagined pleasure, look and feel of the product and happiness in using the product in context (Goodman et al. 2012, p. 23). Product Reaction Cards (PRC) based assessment has been recognised as one of the preferred methods for assessing the perceived desirability of visual designs (Albert & Tullis, 2013, p. 144; Barnum & Palmer, 2010B). PRC consists of a pack of 118 cards with 60% positive and 40% negative or neutral adjectives, from which participants choose the words that reflected their feelings toward their interactive experience with a product. According to Barnum and Palmer (2010B, p. 257), PRC unlocks information regarding the user's sense of the quality of information in a more revealing way than any other tool or techniques they had tried. Moreover, they highlighted specific strengths and values of using PRC complex information systems: as users they are able to express their feeling about their interactive experience to show what they like and dislike.

Accordingly, PRC is an instrument that can be used to get useful insights and perceptions of a product effectively, and PRC-based assessments encourage users to provide a balanced view of both positive and negative feedback. User feedback can be analysed to determine the extent to which the tested product meets user requirements and expectations, and to identify design issues and deficiencies affecting the product.

## 3  Analysis of Product Reaction Cards Assessment

In this research study, the product reaction cards evaluation aimed to gauge the overall acceptability of the product tested by the user participant. Determining the optimal sample size for a user experience research study is important. Based on a large number of studies, Faulkner (2003) summarised that a sample size of 15 users will be able discover minimum of 90% of usability issues and mean of 97.050% usability issues. For user testing experiments, Alroobaea and Mayhew (2014) reported that the optimum sample size of 16±4 users provide much validity. A total number of 32 test users participated in this study. Test users were grouped into two independent categories of 16 users. These two groups of test users were used for both individual product evaluations. Upon the completion of individual product evaluation on CAP and EAP product, users were asked to refer to the product reaction cards checklist and choose all words that best described their interactive experience with the product. Users were then asked to refine their selection and narrow down it to the top five product reaction cards words and rank them in the order of importance (5 being the most important and 1 being the least important). There were 16 product reaction cards evaluations for each of the two products totalling 32 for both products CAP and EAP.

Table 1 and Table 2 show the participant choices in five categories that have been ranked from one to five for both products CAP and EAP. These PRC word choices can be considered as ordered categorical data.

| CAP Product | | | | | |
|---|---|---|---|---|---|
| Participant ID | Ranking5 | Ranking4 | Ranking3 | Ranking2 | Ranking1 |
| u1 | Ambiguous | Poor Quality | Faulty | Rigid | |
| u2 | Poor Quality | Unattractive | Inadequate | | |
| u3 | Ambiguous | Poor Quality | Unattractive | Misleading | Confusing |
| u4 | Ambiguous | Annoying | Frustrating | Poor Quality | Unrefined |
| u5 | Ambiguous | Confusing | Inadequate | Controllable | Simple |
| u6 | Confusing | Unattractive | Ordinary | | |
| u7 | Insecure | Ordinary | Dull | Business-like | Creative |
| u8 | Vague | Confusing | Annoying | Distracting | Inadequate |
| u9 | Confusing | Approachable | Simple | Controllable | Easy to use |
| u10 | Simple | Flexible | Fast | Easy to use | Useful |
| u11 | Effective | Easy to use | Time saving | Simple | Clear |
| u12 | Frustrating | Distracting | Cluttered | Ambiguous | Usable |
| u13 | Useful | Usable | Time saving | Satisfying | Organised |
| u14 | Usable | New | Meaningful | clear | Creative |
| u15 | Annoying | Awkward | Boring | Cluttered | Complex |
| u16 | Hard to use | Confusing | Ambiguous | Annoying | Awkward |

Table 1.  Ranked PRC words for CAP product

| EAP Product | | | | | |
|---|---|---|---|---|---|
| Participant ID | Ranking5 | Ranking4 | Ranking3 | Ranking2 | Ranking1 |
| u1 | Usable | Accessible | Approachable | Ordinary | Inconsistent |
| u2 | Unattractive | Inadequate | Ineffective | Unrefined | Vague |
| u3 | Ambiguous | Confusing | Frustrating | Inadequate | Incomprehensible |
| u4 | Accessible | Easy to use | Efficient | Stable | Responsive |
| u5 | Usable | Clear | Easy to use | Efficient | Satisfying |
| u6 | Useful | Usable | Understandable | Satisfying | Meaningful |
| u7 | Useful | Meaningful | Easy to use | Simple | Organised |
| u8 | Useful | Usable | Straightforward | Simple | Ordinary |
| u9 | Useful | Easy to use | Responsive | Business like | Consistent |
| u10 | Usable | Straightforward | Approachable | Accessible | Relevant |
| u11 | Accessible | Effortless | Ordinary | Business-like | Clear |
| u12 | Useful | Understandable | Clear | Approachable | Accessible |
| u13 | Useful | Effective | Easy to use | Approachable | Simple |
| u14 | Usable | Easy to use | Friendly | Satisfying | Simplistic |
| u15 | Effective | Efficient | Useful | Usable | Clean |
| u16 | Easy to use | Useful | Usable | Clear | Appealing |

Table 2.  Ranked PRC words for EAP product

Each of the PRC word choices was assigned with numbers ranging from one to five (from 'Ranking1 = 1' to 'Ranking5 = 5'). Then the total scores for each PRC word were calculated by adding each assigned value for each PRC word occurrence in all five categories.

Table 3 shows the total score for each attribute for both CAP and EAP product.

| CAP Product | |
|---|---|
| PRC Attribute | Total Score |
| Ambiguous | 25 |
| Confusing | 23 |
| Poor Quality | 15 |
| Annoying | 14 |
| Simple | 11 |
| Unattractive | 11 |
| Usable | 10 |
| Frustrating | 8 |
| Easy to use | 7 |
| Inadequate | 7 |
| Ordinary | 7 |
| Distracting | 6 |
| Time saving | 6 |
| Useful | 6 |
| Awkward | 5 |
| Cluttered | 5 |
| Effective | 5 |
| Hard to use | 5 |
| Insecure | 5 |
| Vague | 5 |
| Approachable | 4 |
| Controllable | 4 |
| Flexible | 4 |
| New | 4 |
| Boring | 3 |
| clear | 3 |
| Dull | 3 |
| Fast | 3 |
| Faulty | 3 |
| Meaningful | 3 |
| Business-like | 2 |
| Creative | 2 |
| Misleading | 2 |
| Rigid | 2 |
| Satisfying | 2 |
| Complex | 1 |
| Organised | 1 |

| EAP Product | |
|---|---|
| PRC Attribute | Total Score |
| Useful | 37 |
| Usable | 33 |
| Easy to use | 26 |
| Accessible | 17 |
| Approachable | 10 |
| Clear | 10 |
| Effective | 9 |
| Efficient | 9 |
| Straightforward | 7 |
| Understandable | 7 |
| Inadequate | 6 |
| Ordinary | 6 |
| Ambiguous | 5 |
| Meaningful | 5 |
| Satisfying | 5 |
| Simple | 5 |
| Unattractive | 5 |
| Business like | 4 |
| Confusing | 4 |
| Effortless | 4 |
| Responsive | 4 |
| Friendly | 3 |
| Frustrating | 3 |
| Ineffective | 3 |
| Stable | 2 |
| Unrefined | 2 |
| Appealing | 1 |
| Clean | 1 |
| Consistent | 1 |
| Incomprehensible | 1 |
| Inconsistent | 1 |
| Organised | 1 |
| Relevant | 1 |
| Simplistic | 1 |
| Vague | 1 |

*Table 3. Total score of each PRC attributes for CAP and EAP product*

In order to make a comparison of both products in terms of PRC attributes, attributes that are common in both products were selected. These common attributes and their corresponding total scores for both products (product 1 and product 2) are shown in Table 4. PRC attributes that are not common to both products were ignored.

| PRC Attributes | CAP Product Total Score | EAP Product Total Score |
| --- | --- | --- |
| Ambiguous | 25 | 5 |
| Confusing | 23 | 4 |
| Frustrating | 8 | 3 |
| Inadequate | 7 | 6 |
| Ordinary | 7 | 6 |
| Simple | 11 | 5 |
| Unattractive | 11 | 5 |
| Unrefined | 1 | 2 |
| Vague | 5 | 1 |
| Approachable | 4 | 10 |
| Business-like | 2 | 4 |
| Clear | 3 | 10 |
| Easy to use | 7 | 26 |
| Effective | 5 | 9 |
| Meaningful | 3 | 5 |
| Organised | 1 | 1 |
| Satisfying | 2 | 5 |
| Usable | 10 | 33 |
| Useful | 6 | 37 |

*Table 4. Total score of PRC attributes that are common for both CAP and EAP products*

Common attributes in Table 4 consist of both positive and negative items. In order to make the product comparison in terms of positive as well as negative PRC attributes, a further separation was made to distinguish the positive and negative PRC attributes from the common attributes in Table 4. Table 5 shows total scores for common positive and negative PRC attributes of product 1 and product 2 separately.

| Positive PRC Attributes | CAP Product Total Score | EAP Product Total Score |
| --- | --- | --- |
| Approachable | 4 | 10 |
| Business-like | 2 | 4 |
| Clear | 3 | 10 |
| Easy to use | 7 | 26 |
| Effective | 5 | 9 |
| Meaningful | 3 | 5 |
| Organised | 1 | 1 |
| Satisfying | 2 | 5 |
| Usable | 10 | 33 |
| Useful | 6 | 37 |

| Negative PRC Attributes | CAP Product Total Score | EAP Product Total Score |
| --- | --- | --- |
| Ambiguous | 25 | 5 |
| Confusing | 23 | 4 |
| Frustrating | 8 | 3 |
| Inadequate | 7 | 6 |
| Ordinary | 7 | 6 |
| Simple | 11 | 5 |
| Unattractive | 11 | 5 |
| Unrefined | 1 | 2 |
| Vague | 5 | 1 |

*Table 5. Common positive and negative PRC attributes for both products*

## 4 Surface Measure of Overall Performance

The radar chart approach based Surface Measure of Overall Performance (SMOP) has been recognised as a technique that obtains the overall performance of a system (Mosley & Mayer, 1998; Schmid et al., 1999; Schütz et al., 1998; Behringer et al., 2005). SMOP is given by the surface area of a radar chart formed by the joined lines of performance indicators represented in each dimension (radial line) of the chart.

Schütz et al. (1998, p. 39) have highlighted the main goals of using a radar-based SMOP approach:

- Visualization of interrelated performance measures through standardized scales.

- Presentation of an effective and revealing description of selective performance dimensions in just one synthetic indicator (using the surface of the radar chart to the illustration of the performance of the system).

- The change in the overall performance between two points of intervals can be analysed.
- The shape of the radar chart, as well as the overall performance measures, can be used for comparisons of systems.

The mathematical formula for calculating SMOP for four axes:

SMOP = (P1*P2)+(P2*P3)+(P3*P4)+(P4*P5)+(P5*P6)+....+(Pn*P1)) * sin90/2

The mathematical formula for calculating SMOP for more or less than four axes:

SMOP = (P1*P2)+(P2*P3)+(P3*P4)+(P4*P5)+(P5*P6)+....+(Pn*P1)) * sin(360/n)/2

Where P = data point on the performance indicator and n = total number of data points.

In SMOP, the maximum value that can be assigned to any radial line is '1'. With reference to Table 5 for positive PRC attributes, 'useful' can be found as the attribute having the highest total score with the value '37'. For SMOP calculations, this highest value 37 is weighted as the calculated highest score '1' and all other total scores are converted as a fraction of '37' to get 'calculated Scores'. Following a similar approach, 'ambiguous' was shown with the highest score among PRC negative attributes with the value 25, and all total scores were expressed as a fraction of 25 to derive 'calculated scores'. For both product 1 and product 2, the calculated scores of an attribute for SMOP calculations are shown in Table 6.

| Positive PRC Attributes | CAP Product Calculated Score | EAP Product Calculated Score |
|---|---|---|
| Approachable | 0.1 | 0.27 |
| Business-like | 0.05 | 0.1 |
| Clear | 0.08 | 0.27 |
| Easy to use | 0.18 | 0.7 |
| Effective | 0.13 | 0.24 |
| Meaningful | 0.08 | 0.13 |
| Organised | 0.02 | 0.02 |
| Satisfying | 0.05 | 0.13 |
| Usable | 0.27 | 0.89 |
| Useful | 0.16 | 1 |

| Negative PRC Attributes | CAP Product Calculated Score | EAP Product Calculated Score |
|---|---|---|
| Ambiguous | 1 | 0.2 |
| Confusing | 0.92 | 0.16 |
| Frustrating | 0.32 | 0.12 |
| Inadequate | 0.28 | 0.24 |
| Ordinary | 0.28 | 0.24 |
| Simple | 0.44 | 0.2 |
| Unattractive | 0.44 | 0.2 |
| Unrefined | 0.04 | 0.08 |
| Vague | 0.2 | 0.04 |

*Table 6. Calculated CAP and EAP attribute values for SMOP Radar Charts*

As shown in Figure 2 and 3, two radar charts were developed from the data of Table 6, representing the performance measure of the corresponding PRC attributes on each radial axis. Figure 1 shows the radar chart for positive PRC attributes.

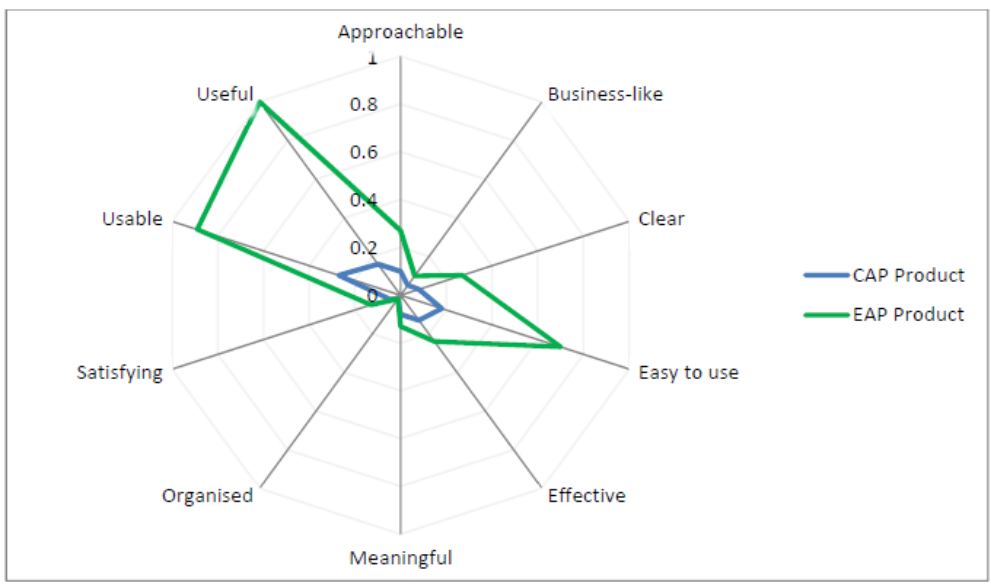

*Figure 2: Radar chart for positive PRC attributes*

The positive PRC attributes represented in Figure 1 are: approachable, business-like, clear, easy to use, effective, meaningful, organise, satisfying, usable and useful. The two overlaying radar charts in Figure 1 depicts the relative system performance of the CAP product and the EAP product in terms of positive PRC attributes. The highest performing three positive PRC attributes are:

-       Useful (performance score = 1)
-       Usable (performance score = 0.89)
-       Easy to use (performance score = 0.7).

Figure 2 illustrates the comparative positive performance between CAP and EAP products. As evident in Figure 2, the EAP product demonstrates a higher level of performance, representing a larger surface area, and the CAP product has a much lower performance than the EAP product, with a significantly smaller surface area of the radar chart.

Figure 3 shows the radar chart for negative PRC attributes.

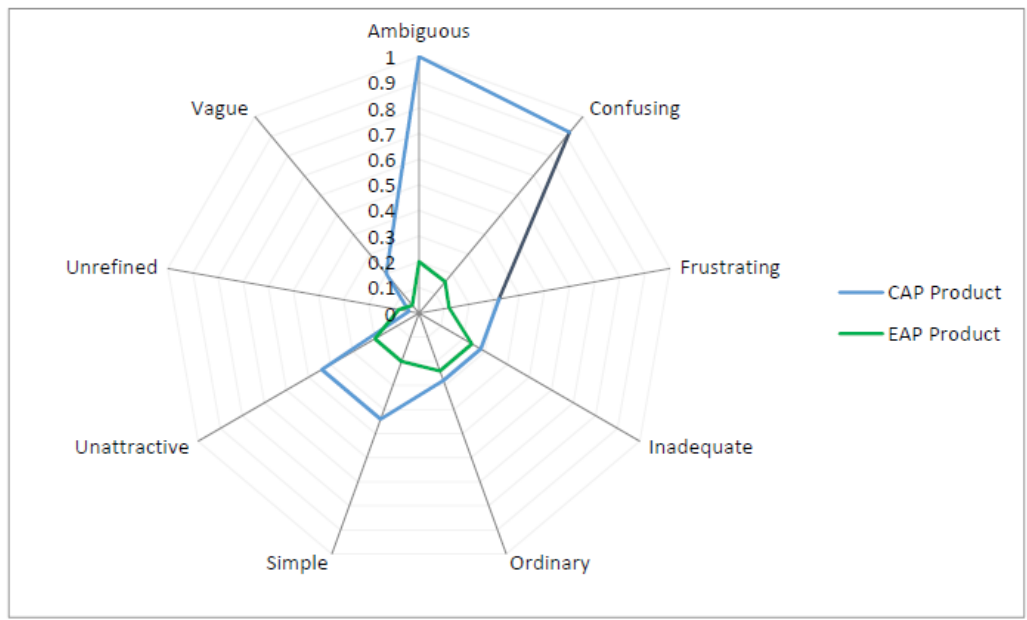

*Figure 3: Radar chart for negative PRC attributes*

The negative PRC attributes represented in Figure 3 are Ambiguous, Confusing, Frustrating, Inadequate, Ordinary, Simple, Unattractive, Unrefined, and Vague. The two overlaying radar charts in Figure 3 depicts the relative system performance of the CAP and EAP products in terms of negative PRC attributes. The highest performing three negative PRC attributes are:

-       Ambiguous (performance score = 1)
-       Confusing (performance score = 0.92)
-       Simple and Unattractive (performance score = 0.44).

Unlike Figure 2, Figure 3 illustrates the comparative negative performance between CAP and EAP products. Hence, the radar chart with the larger surface area corresponds to the product with poor performance, and the radar chart with smaller surface area corresponds to the product with 'less poor performance'. As evident in Figure 3, the CAP product demonstrates a higher level of poor performance representing a larger surface area, and the EAP product performs much better than the CAP product with a significantly smaller surface area of the radar chart.

### 4.1　SMOP calculations – positive PRC attributes in CAP product

SMOP for positive PRC attributes in CAP product is calculated applying the formula:

SMOP (CAP) = (P1*P2)+(P2*P3)+(P3*P4)+(P4*P5)+(P5*P6)+....+(Pn*P1)) * sin(360/n)/2

=(0.1*0.05)+(0.05*0.08)+(0.08*0.18)+(0.18*0.13)+(0.13*0.08)+(0.08*0.02)+(0.02*0.05)+(0.05*0.27)+(0.27*0.16)+(0.16*0.01)*Sin(360/10)*0.5 = **0.038**

### 4.2　SMOP calculations – positive PRC attributes in EAP product

SMOP for positive PRC attributes in EAP product is calculated applying the formula:

SMOP (EAP) = (P1*P2)+(P2*P3)+(P3*P4)+(P4*P5)+(P5*P6)+....+(Pn*P1)) * sin(360/n)/2

=(0.27*0.1)+(0.1*0.27)+(0.27*0.7)+(0.7*0.24)+(0.24*0.13)+(0.13*0.02)+(0.02*0.13)+(0.13*0.89)+(0.89*1.0)+(1.0*0.27)*Sin(360/10)*0.5 = **0.504**

### 4.3　SMOP calculations – negative PRC attributes in CAP product

SMOP for negative PRC attributes in CAP product is calculated applying the formula:

SMOP (CAP) = (P1*P2)+(P2*P3)+(P3*P4)+(P4*P5)+(P5*P6)+....+(Pn*P1)) * sin(360/n)/2

=(1.0*0.92)+(0.92*0.32)+(0.32*0.28)+(0.28*0.28)+(0.28*0.44)+(0.44*0.44)+(0.44*0.04)+(0.04*0.02)+(0.02*1.0)*Sin(360/9)*0.5 = **0.617**

### 4.4　SMOP calculations – negative PRC attributes in EAP product

SMOP for negative PRC attributes in EAP product is calculated applying the formula:

SMOP (EAP) = (P1*P2)+(P2*P3)+(P3*P4)+(P4*P5)+(P5*P6)+....+(Pn*P1)) * sin(360/n)/2

=(0.02*0.16)+(0.16*0.12)+(0.12*0.24)+(0.24*0.24)+(0.24*0.20)+(0.20*0.20)+(0.20*0.08)+(0.08*0.04)+(0.04*0.02)*Sin(360/9)*0.5= **0.081**

### 4.5　Summary of SMOP calculations

Table 7 shows the summary of results of SMOP calculations.

| PRC Positive Attributes - Summary of Results | | |
|---|---|---|
| **Product** | **SWOP Value** | **Remarks** |
| CAP | 0.038 | In terms of PRC positive attributes, EAP product shows a significantly more effective performance with a much higher SWOP score of 0.504. |
| EAP | 0.504 | |

| PRC Negative Attributes - Summary of Results | | |
|---|---|---|
| **Product** | **SWOP Value** | **Remarks** |
| CAP | 0.617 | In terms of PRC negative attributes, CAP product shows a significantly poorer performance with a much higher SWOP score of 0.617. |
| EAP | 0.081 | |

*Table 7.　Summary of results of SMOP calculations*

Results of the PRC analysis represented by the two radar charts (see Figure 4 and 5 which shows both radar charts in Figure 2 and 3 for comparison) show that the EAP product is significantly more 'effective' than the CAP product.

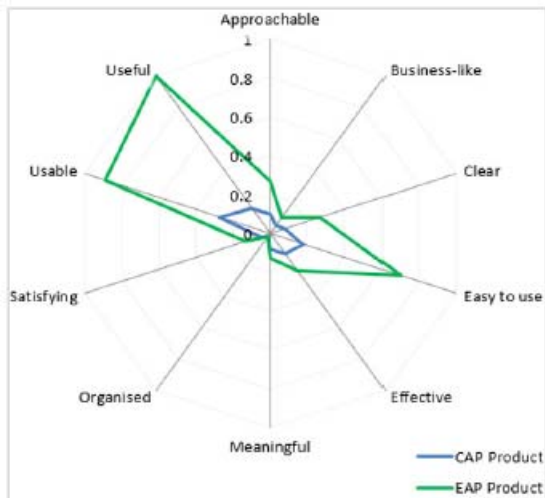 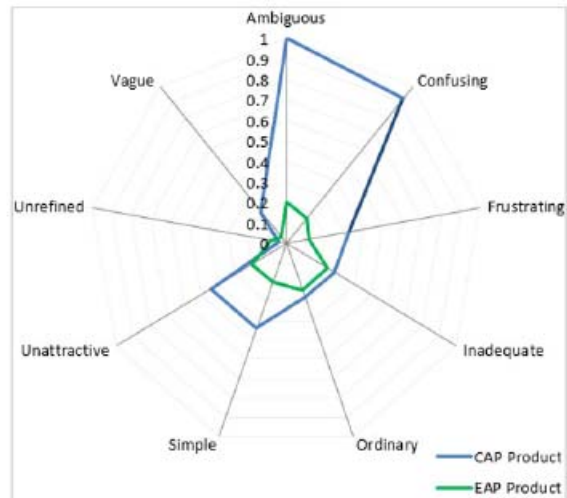

*Figure 4: Analysis: PRC positive attributes*　　　*Figure 5: Analysis: PRC negative attributes*

The overall conclusion that can be drawn from the results of product reaction cards evaluation is that the EAP product has greater positive PRC attributes and fewer negative *PRC attributes.*

## 5 Conclusions

This paper presents a quantitative data analysis of a comparative desirability assessment of two software products developed based on two conceptually different design approaches. The data collection was conducted using 118 item Microsoft Product Reaction Cards (PRC) tool, followed by data analysis based on the Surface Measure of Overall Performance (SMOP) approach. The main contribution of this research comes from the integrated method that incorporated product reaction cards and surface measure of performance for data collection and data analysis. The product reaction cards evaluation introduced two novel techniques for data analysis: a novel weighting system to scale the data for analysis, and the incorporation of the Surface Measure of Overall Performance as a technique of data analysis to derive conclusive results. The results of this study suggest that the incorporation of SMOP as an approach for PRC data analysis is effective in deriving conclusive evidence of desirability in user experience.

## Copyright